\documentclass[namedreferences]{solarphysics}
\usepackage[optionalrh]{spr-sola-addons} 
\usepackage{graphicx,rotating}        
\usepackage{amssymb}        
\usepackage{color}           
\usepackage{url}             

\chardef\us=`\_

\begin{document}

\begin{article}

\begin{opening}

\title{Fragile detection of solar g modes by Fossat \textit{et al.}  }


\author[addressref={aff1},corref,email={schunker@mps.mpg.de}]{\inits{H.}\fnm{Hannah}~\lnm{Schunker}}
\author[addressref=aff1]{\inits{J.}\fnm{Jesper}~\lnm{Schou}}
\author[addressref=aff1]{\inits{P.}\fnm{Patrick}~\lnm{Gaulme}}
\author[addressref={aff1,aff2}]{\inits{L.}\fnm{Laurent}~\lnm{Gizon}}

\address[id=aff1]{Max-Planck-Institut f\"ur Sonnensystemforschung, Justus-von-Liebig-Weg 3, 37077 G\"ottingen}
\address[id=aff2]{Georg-August-Universit\"at G\"ottingen, Institut f\"ur Astrophysik, Friedrich-Hund-Platz 1, 37077 G\"ottingen}

\runningauthor{Schunker et al.}
\runningtitle{Fragile detection of solar g modes}

\begin{abstract}
The internal gravity modes of the Sun are notoriously difficult to detect, and the claimed detection of gravity modes presented in \citet{Fossatetal2017} is thus very exciting. 
Given the importance of these modes for understanding solar structure and dynamics, the results must be robust.
While \citet{Fossatetal2017} described their method and parameter choices in detail, the sensitivity of their results to several parameters were not presented. 
Therefore, we test the sensitivity to a selection of them.
The most concerning result is that the detection vanishes when we adjust the start time of the 16.5~year velocity time series by a few hours. 
We conclude that this reported detection of gravity modes is extremely fragile and should be treated with utmost caution.
\end{abstract}
\end{opening}


\section{Introduction}\label{sect:intro}
      
Despite the revelations about the solar internal structure and dynamics from helioseismology in the past thirty years, the deep core of the Sun has remained invisible.
This is because the most easily observed pressure modes (p modes) are predominantly sensitive to the near-surface layers of the Sun. Gravity modes (g modes) that probe the core of the Sun, are evanescent in the convection zone and have small amplitudes at the surface making them difficult to detect (e.g. \opencite{Appourchauxetal2010}).
There have been prior claims of g-mode detection, (e.g. \opencite{Garciaetal2007}), but to our knowledge the results have not been independently reproduced, and remain controversial. Refreshingly, \citet{Fossatetal2017} provide their data publicly and describe their method sufficiently well for us to qualitatively reproduce their results.

\citet{Fossatetal2017} present a method based on the principle that g modes perturb the solar core, changing the round-trip travel-time (the time taken to travel to the other side of the Sun and back) of the sound waves (p modes). 
They measure perturbations to the large separation (equivalent to the round-trip travel-time) between pairs of low-frequency p modes with even ($\ell=0, 2$) and odd ($\ell=1, 3$) harmonic degrees. 
The large separation is not as susceptible to convective noise and surface effects as individual mode frequencies, and is sensitive to the mean density of the Sun.

Applying their method to the long-term velocity time series from the Global Oscillations at Low Frequencies (GOLF) instrument \citep{GOLF1995}, \citet{Fossatetal2017} inferred that the solar core rotates about 3.8 times faster than the envelope.
Given the potential impact of this detection on solar structure and dynamics, it is important that these results can be independently reproduced and tested.
A robust detection will also impact the theory of stellar evolution,
and the method could potentially be applied to stellar observations of Sun-like stars, for example from PLATO \citep{PLATO2014} observations.

Although \citet{Fossatetal2017} described their analysis method well enough to be qualitatively reproduced, they did not present a quantitative description of the sensitivity to the parameters of the method.

In this paper we first present our independent reproduction of the measurement of the rotational splitting of the g modes (Sect.~\ref{sect:ogresults}).
We then examine the sensitivity of the significance of the detection to four parameters in the analysis method: the method used to measure the round-trip travel-time in Sect.~\ref{sect:fits}, the smoothing of the power spectrum of the round-trip travel-time in Sect.~\ref{sect:smooth}, the cadence of the round-trip travel-time measurements in    Sect.~\ref{sect:seglen}, and the start-time of the GOLF time series in Sect.~\ref{sect:starttime}.
We conclude in Sect.~\ref{sect:conc}.


\section{Reproduction of g mode detection} \label{sect:ogresults}

By  following the analysis method described in \citet{Fossatetal2017} we were able to qualitatively reproduce the results of Fig.~10 in that paper. 
However our results are slightly different, possibly due to differences in the input data or because some parts of the algorithm were not clearly described. Here we describe our procedure in relation to that described in \citet{Fossatetal2017}.

We downloaded the 16.5~year-long time series of the solar global velocity observations from the GOLF instrument provided at \url{https://www.ias.u-psud.fr/golf/templates/access.html} and divided the data into 36130, 8~hour-long segments with a 4~hr cadence.  
 
For each valid segment we first padded the data out to $10^6$~seconds before computing the power spectrum. 
We filtered this power spectrum for the frequency band between 2.32 and 3.74~mHz,
and translated the beginning of this band to the zeroth frequency. 
We subtracted the mean of the power spectrum and zero-padded out to 125~mHz.
We then computed the power spectrum of this padded, frequency-filtered power spectrum, to get the temporal power spectrum (which is the envelope of the autocorrelation of the GOLF time series in the selected frequency range).
We did a least-squares fit of a quadratic function to the prominent peak in the range between $14000$~s and $15600$~s. The time of the maximum of the quadratic is the round-trip travel-time of the p modes.
At this point, we have a time series of the round-trip travel-time of the p modes at a 4~hr cadence. 
If a segment had a duty-cycle less than 25\% we set the measurement of the round-trip travel-time to zero.

The condition that the segments have a minimum 25\% duty-cycle  was a subjective choice on our behalf because the value was not specifically stated in \citet{Fossatetal2017}, and results in 1.5\% more round-trip travel-time measurements than \citet{Fossatetal2017}.
There are 34775 non-zero values with a root-mean-square (rms) uncertainty of 54~s, compared to Fig.~6 in \citet{Fossatetal2017}, who have 34261 non-zero values with an rms of 52~s.

We then subtracted the mean value (14806~s, compared to \citet{Fossatetal2017} who gets 14807~s) to get the round-trip travel-time perturbation as a function of time with a 4~hr cadence.
We set any absolute value larger than  240~s to zero.
We then computed the power spectrum of these round-trip travel-time perturbations, and convolved it with a six-pixel-wide box-car window (11.5~nHz) to smooth the power spectrum, and calculated it's autocorrelation up to 3500~nHz. We note that \citet{Fossatetal2017} used a seven-pixel-wide window with weighting $[0.5, 1 ,1 ,1 ,1 ,1 ,0.5]$ (private communication, E.~Fossat). This makes little difference to the results.

\begin{figure}
\includegraphics[width=0.8\textwidth]{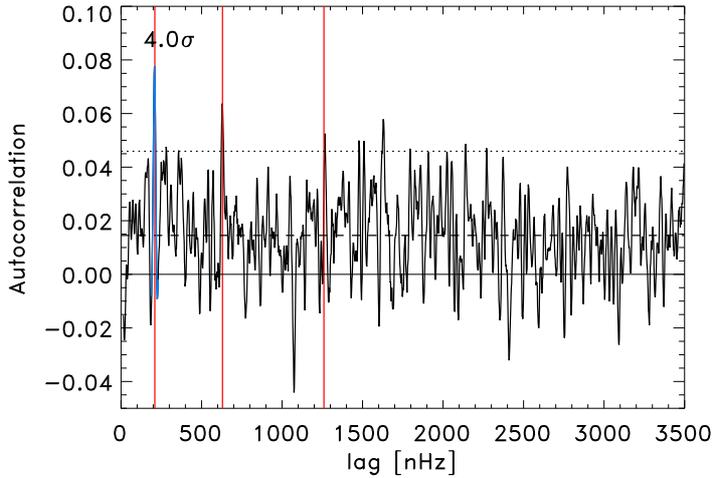}
\caption{Autocorrelation of the power spectrum. This figure is directly comparable to Fig.~10 in Fossat \textit{et al.} (2017). The vertical red lines indicate the frequency splittings detected by Fossat \textit{et al.} (2017) at 210, 630 and 1260~nHz. The horizontal dashed line is the mean of the autocorrelation and the dotted line represents two standard deviations. The blue curve shows the range within which we compute the significance. 
}
\label{fig:perturb}
\end{figure}

Figure~\ref{fig:perturb} shows the result of our analysis, which can be qualitatively compared to Fig.~10 of \citet{Fossatetal2017}.
The peaks at 210~nHz, 630~nHz and 1260~nHz are claimed by \citet{Fossatetal2017} to represent the splittings of the g modes caused by rotation in the core.
Our highest peak at 210~nHz has a significance of $4.0~\sigma$ compared to $4.7\sigma$ in \citet{Fossatetal2017}.

We defined the significance as the maximum value of the autocorrelation in a certain range of frequency lag, relative to the standard deviation of the autocorrelation in the full range of frequency lag. 
We defined the maximum of the first peak to lie within the range 189-231~nHz (see the blue part of the curve in Figs.~\ref{fig:perturb}).

Here, we simply aimed to first qualitatively reproduce the results, and in the following sections we will analyse the sensitivity to some parameters in the analysis method.

Initially, we did not interpret the description of ``smoothed over 6 bins" and ``6-bin smoothing" of the power spectrum correctly and had difficulty acquiring the large significance of the main peak. Through private communication with D.~Salabert and T.~Corbard we established that this phrase means a convolution of a six-pixel-wide window with the power spectrum.

We found that the significance of our main peak could increase by relaxing the criteria that each 8~hour segment of data has a duty-cycle of at least 25\%. However, we chose to keep a well defined value here rather than try to tune any of the criterion to increase the significance.

\section{Sensitivity to the measurement of the round-trip travel-time}\label{sect:fits}

\citet{Fossatetal2017} claim that ``the (second-order) polynomial fit made on a range of $\pm800$~s around $14~800$~s minimizes the scatter of $T$", (where $T$ is the round-trip travel-time) and that this fit is ``not at all the best fit on the peak profile, but it is the least noisy estimate of the peak centroid.". 

\begin{figure}
\includegraphics[width=0.9\textwidth]{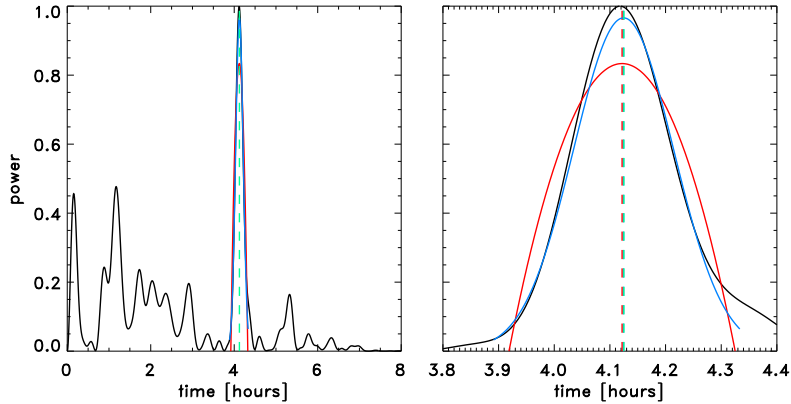}
\caption{Left: Example temporal power spectrum of the fourth 8 hour-long segment showing the round-trip travel-time peak. This can be directly compared to Fig.~4 in  Fossat \textit{et al.} (2017). The red curve is the quadratic fit as was done in Fossat \textit{et al.} (2017), the blue curve is a Gaussian fit. The vertical dashed lines show the location of the measured round-trip travel-time for each measurement method. The location of the maxima of the quadratic (red, 14838.7~s), the Gaussian (blue 14848.4~s)  and centroid (green, 14849.9~s) are barely distinguishable on this scale. Right: the same as in the left panel, but restricted to the time range used to measure the round-trip travel-time. 
}
\label{fig:fits}
\end{figure}

Naively, it looks like a Gaussian would be a good fit to measure the peak of this curve (see Fig.~\ref{fig:fits}), however the centroid may be the key parameter. 
Therefore we compare the least squares fit of the quadratic function used in \inlinecite{Fossatetal2017}, 
$q(t)=at^2 + bt + c$ (where $t$ is time), 
with a non-linear least squares fit  of a three parameter Gaussian function, 
$g(t)= A e^{\left( \frac{(t-B)^2}{2C^2} \right)}$;
and a direct measurement of the centroid   
$\langle c \rangle = \frac{ \int_{t_1}^{t_2} t \, T(t) dt  }{ \int_{t_1}^{t_2}  T(t) dt }$, 
where $t_1=14000$~s, $t_2=15600$~s and we evaluated the integrals discretely using the trapezoidal rule.
Figure~\ref{fig:fits} shows an example of the different fits to the round-trip travel-time peak.

The root-mean-square of the round-trip travel-time perturbations (e.g. similar to Fig.~\ref{fig:perturb}) is    
53~s for the quadratic fit, 
72~s for the Gaussian fit and 
64~s for the centroid measurement, 
compared to 52~s in \citet{Fossatetal2017}.

\begin{figure}
\includegraphics[width=0.9\textwidth]{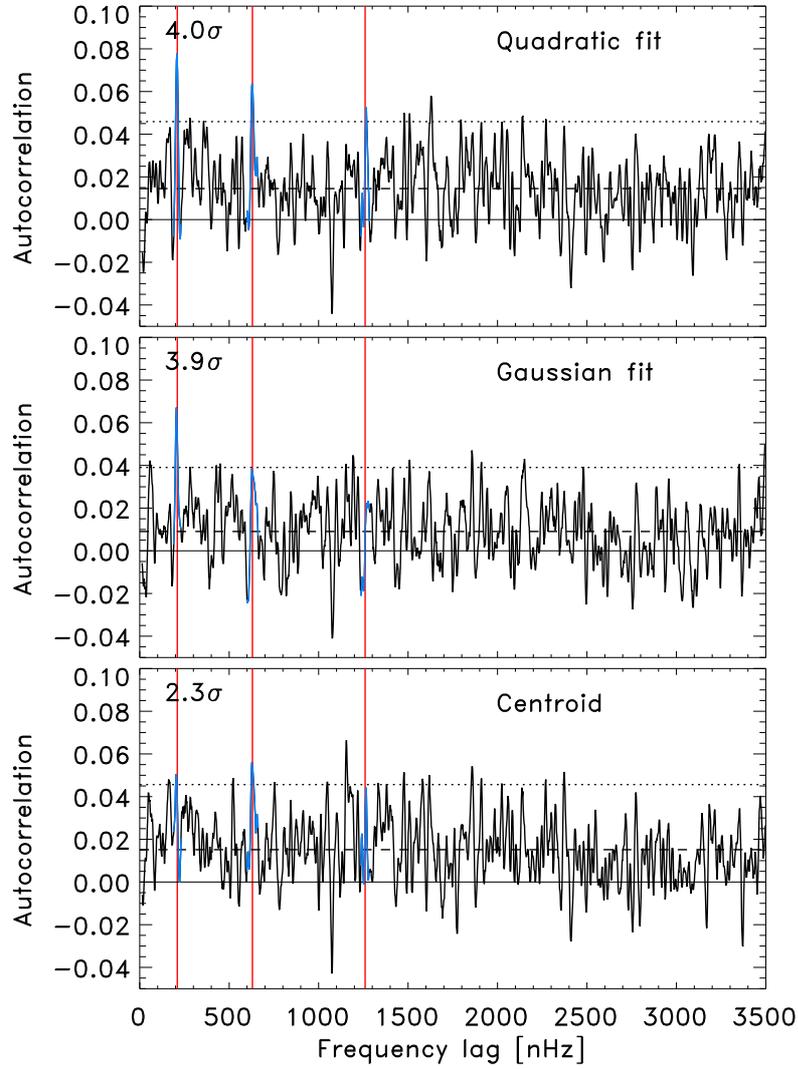}
\caption{The autocorrelation function for the three different measurements of the round-trip travel-time. The top panel is the same as Fig.~\ref{fig:perturb}, the middle panel is the auto-correlation for the Gaussian fits and the bottom panel is for the centroid measurement. The vertical red lines indicate the expected frequency splittings as in Fossat \textit{et al.} (2017). The horizontal dashed line is the mean of the autocorrelation and the dotted line represents two standard deviations. The blue curve shows the range over which we find the maximum to compute the significance of the first peak.}
\label{fig:acfits}
\end{figure}

This quantitatively shows that the quadratic fit does indeed have the lowest noise compared to the two other methods of measuring the round-trip travel-time.
The Pearson correlation coefficient for the time series of the round-trip travel-time measurements
quadratic and Gaussian measurements is 0.89, and the 
quadratic and centroid measurement is 0.93, 
showing that the different methods do not measure something wildly different. 
The significance of the first peak in the autocorrelation function for the Gaussian is similar to the quadratic, but the second peak is not as significant and the third peak is not clear at all (Fig.~\ref{fig:acfits}). The first peak is much less significant in the centroid measurements than in the other two cases. 

\section{Sensitivity to the smoothing of the power spectrum}\label{sect:smooth}

\begin{figure}
\includegraphics[width=0.9\textwidth]{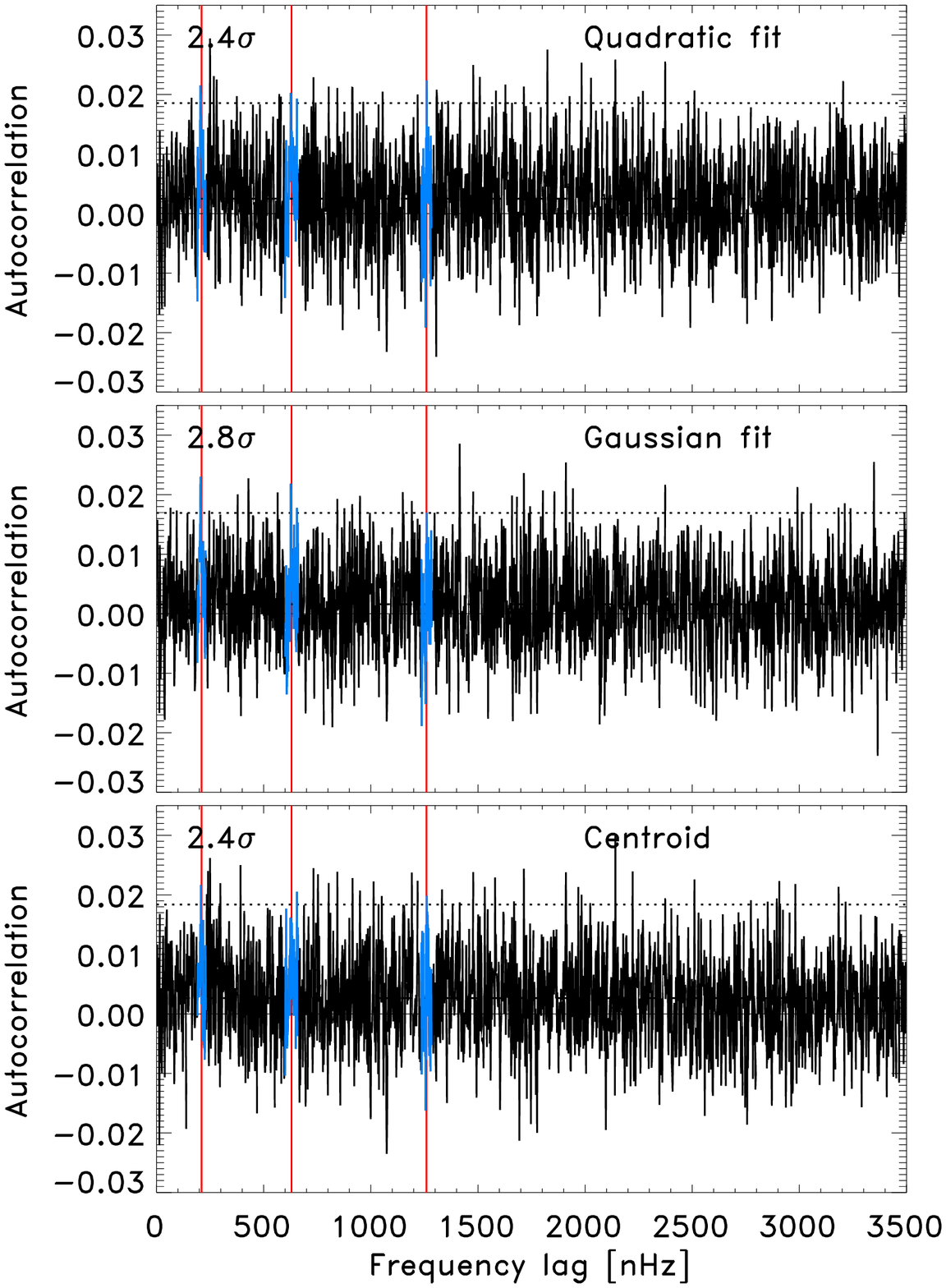}
\caption{The autocorrelation of the power spectrum without any smoothing.  The vertical red lines indicate the expected frequency splittings as in Fossat \textit{et al.} (2017). The horizontal dashed line is the mean of the autocorrelation and the dotted line represents two standard deviations. The blue curve indicates the range over which we compute the significance of the peaks.  
}
\label{fig:acfitsnosmooth}
\end{figure}

\begin{figure}
\includegraphics[width=0.9\textwidth]{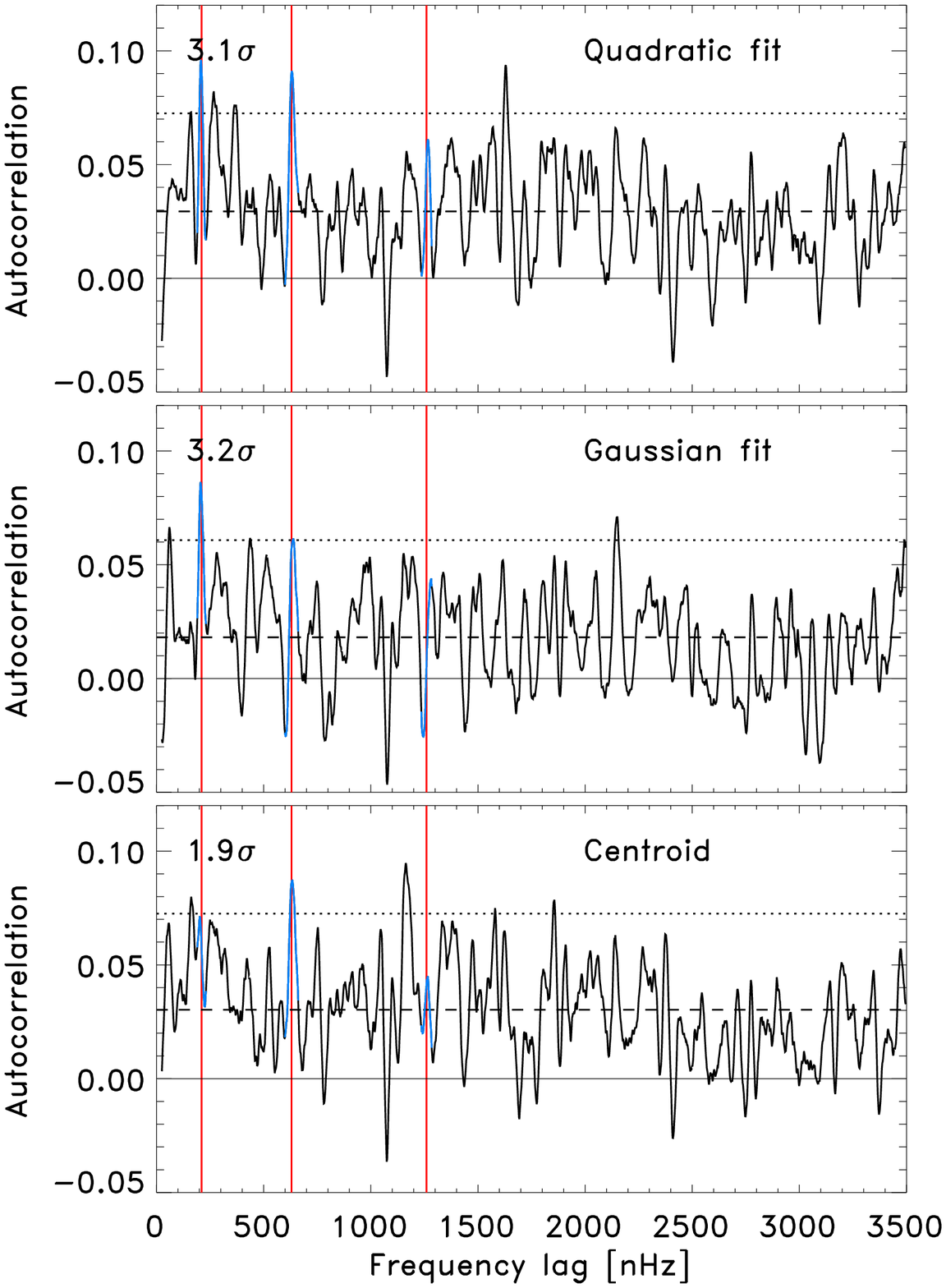}
\caption{The autocorrelation of the power spectrum smoothed by 12~pixels (23~nHz). The vertical red lines indicate the frequencies identified by Fossat \textit{et al.} (2017) as the expected splittings. The horizontal dashed line is the mean of the auto-correlation and the dotted line represents two standard deviations. The blue curve indicates the range over which we find the maximum to compute the significance of the peaks.  }
\label{fig:acfitssmooth12}
\end{figure}

\citet{Fossatetal2017} states that the significance of the peaks is maximised by convolving the power spectrum of the round-trip travel-times with a six-pixel window (11.5~nHz, 1~pixel corresponds to a frequency bin of 1.92~nHz) to smooth the data.
They also claim that the smoothing accounts for the imprecision of the observed mode frequencies in the power spectra.
We computed the autocorrelation for power spectra convolved with box-car windows up to twelve~pixels wide.
Figure~\ref{fig:acfitsnosmooth} shows the autocorrelation of the power spectrum without smoothing, and  Figure~\ref{fig:acfitssmooth12} shows the autocorrelation of the power spectrum smoothed by a twelve-pixel-wide  (23~nHz) box-car window.

\begin{figure}
\includegraphics[width=0.95\textwidth]{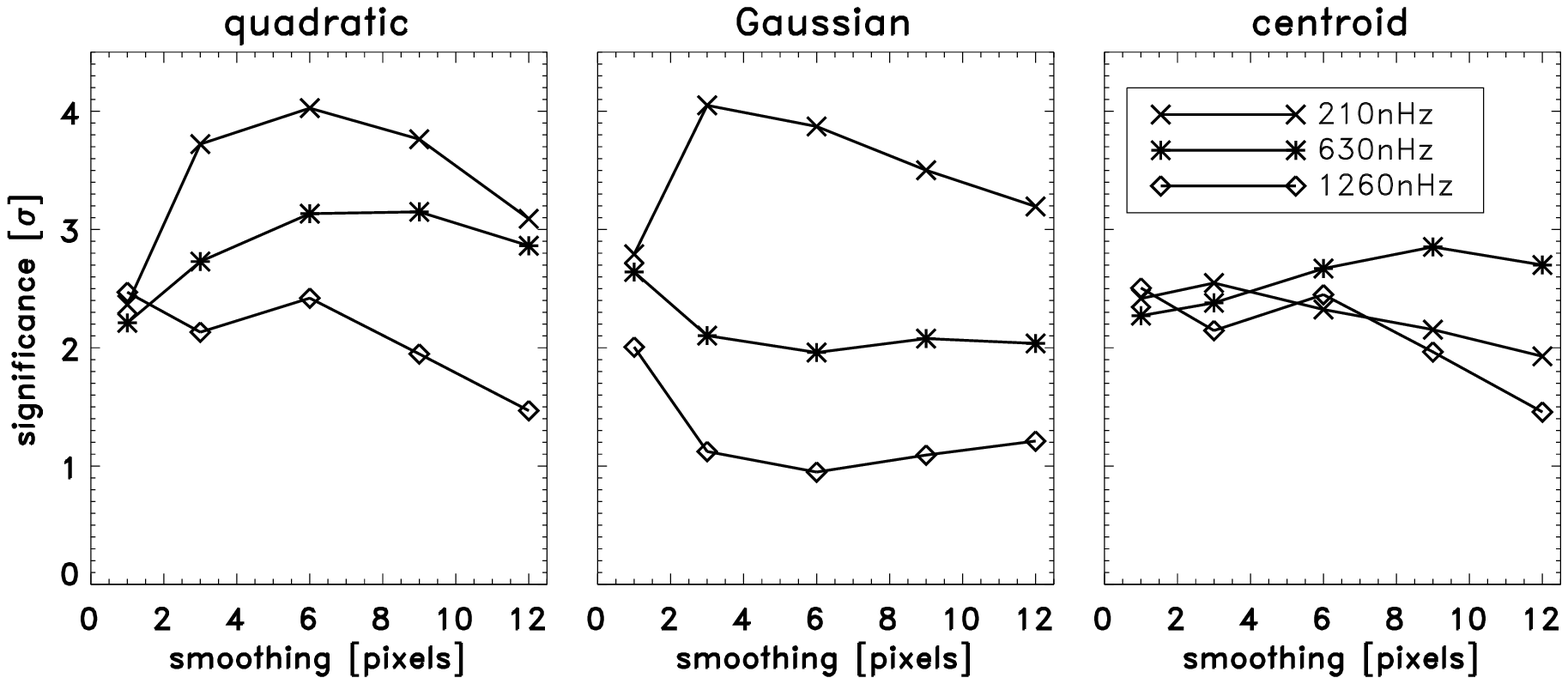}
\caption{Significance of each of the peaks (as indicated by the blue curves in Figs.~3 and 4) as a function of the width of the smoothing window. The quadratic fit with six~pixel-wide smoothing window gives the highest significance for all peaks, in agreement with Fossat \textit{et al.}, 2017.}
\label{fig:acfitsmoothing}
\end{figure}

We computed the significance of the peaks in the autocorrelation by measuring the maximum value in the range 189-231~nHz for the first peak, 598.5-661.5~nHz for the second peak and 1234.8-1285.2~nHz for the third peak (see, for example, the blue parts of the autocorrelation curve in Fig.~\ref{fig:acfits}).
Figure~\ref{fig:acfitsmoothing} shows the significance of each of the three peaks for different sizes of the smoothing window and for the different methods of measuring the round-trip travel-time. 
We quantitatively show that the significance of the first and second peaks is a maximum for a six-pixel-wide smoothing window for the quadratic case as claimed by \citet{Fossatetal2017}.

\section{Sensitivity to start-time of the GOLF time series}\label{sect:starttime}

The GOLF velocity time series is 16.5~years long with an 80~s cadence.
We changed the start-time of the data by removing different amounts of data from the beginning of the time series to test the stability of the results.

\begin{figure}
\includegraphics[width=0.95\textwidth]{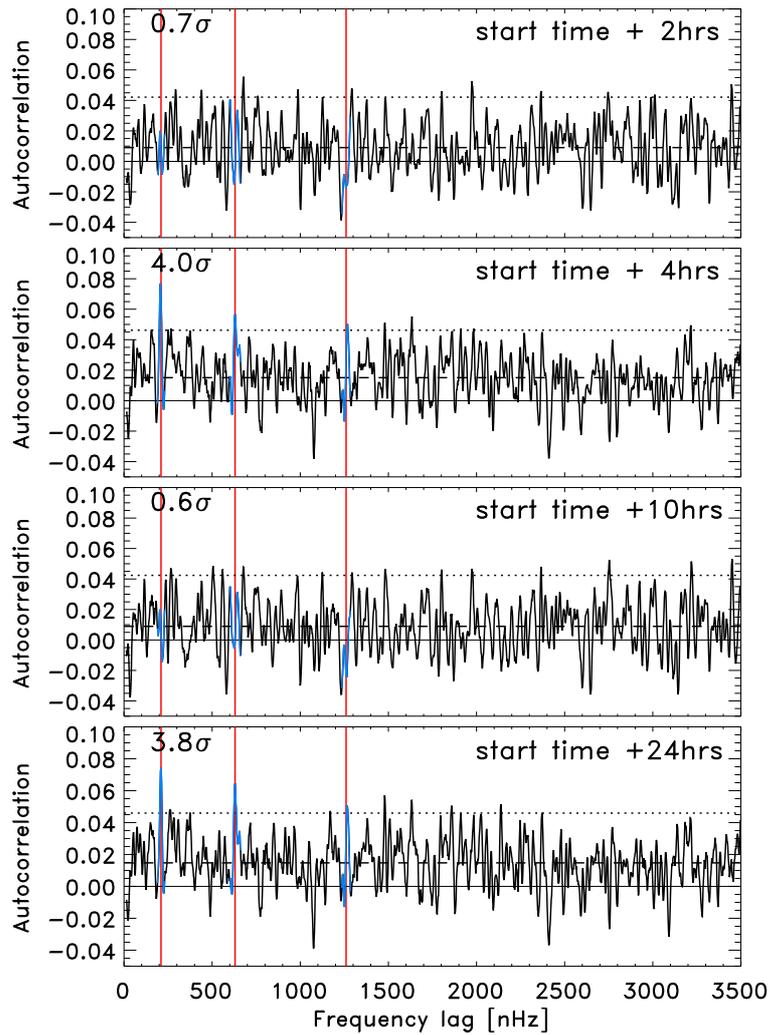}
\caption{The autocorrelation function for start-times offset by 2~hrs (top), 4~hrs, 10~hrs and 24~hrs (bottom) relative to the original start-time. 
The blue curve indicates the range within which we find the maxima to compute the significance of the peak. 
The vertical red lines indicate where Fossat \textit{et al.} (2017) identified the original three g-mode\ peaks. 
Peaks at the location of the purported g modes\ are only evident at multiples of four hours.}
\label{fig:acchop}
\end{figure}

Figure~\ref{fig:acchop} shows the autocorrelation function for four different start times. When we removed 2~hrs and 10~hrs there are no significant peaks. On the other hand, removing whole segments (e.g. 4 or 24 hours), does not have a large effect.
That the significance is so sensitive to shifting the start time by a couple of hours out of 16.5 years clearly shows that the result is fragile.
 
\begin{figure}
\includegraphics[width=0.9\textwidth]{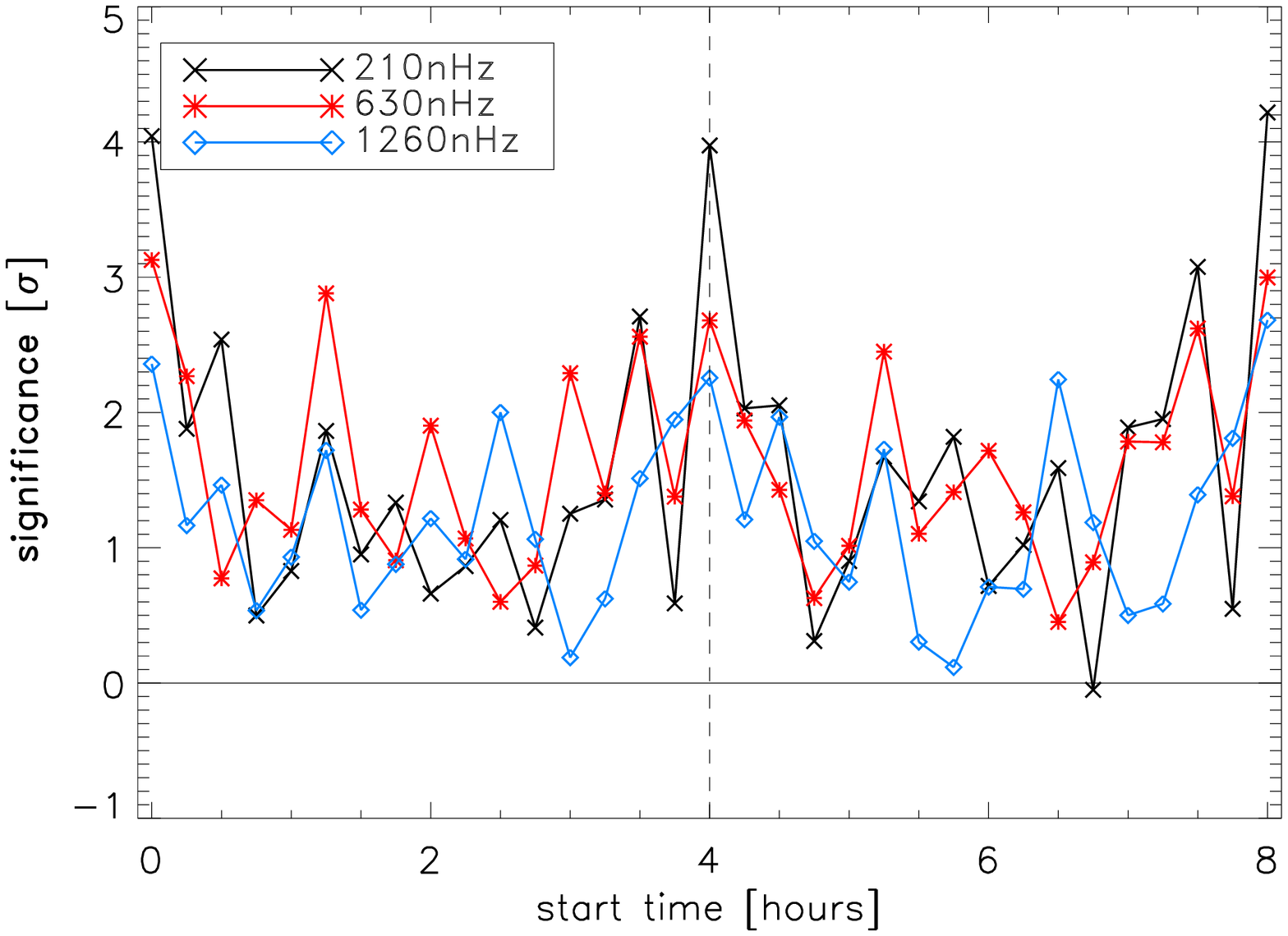}
\caption{Significance of the three g-mode\ peaks as a function of start-time relative to the start-time of the original GOLF time series. 
The black crosses are the significance of the maximum value of the autocorrelation near the first peak, the red stars are the significance near the second peak, and the blue diamonds are the significance near the third peak. 
The first points at zero start-time correspond to the three peaks in Fig.~\ref{fig:perturb}. 
}
\label{fig:acchopfine}
\end{figure}

To further investigate this, Fig.~\ref{fig:acchopfine} shows the significance of the three peaks as a function of the start time. This shows a 4~hr oscillation period, equal to the cadence, confirming the fragility of the results.

\section{Sensitivity to cadence of round-trip travel-time measurement}\label{sect:seglen}

\citet{Fossatetal2017} measured round-trip travel-times at a cadence of half (4~hrs) the segment length (8~hrs). These seem like reasonable choices, but were not justified.
We repeated the analysis using different cadences (while keeping the segment length equal to twice the cadence) and scaling the cut-off value of 240~s by the relative rms of the non-zero round-trip travel-time.
Longer (shorter) segment lengths makes the round-trip travel-time peak narrower (wider), and less (more) noisy. 

\begin{figure}
\includegraphics[width=0.9\textwidth]{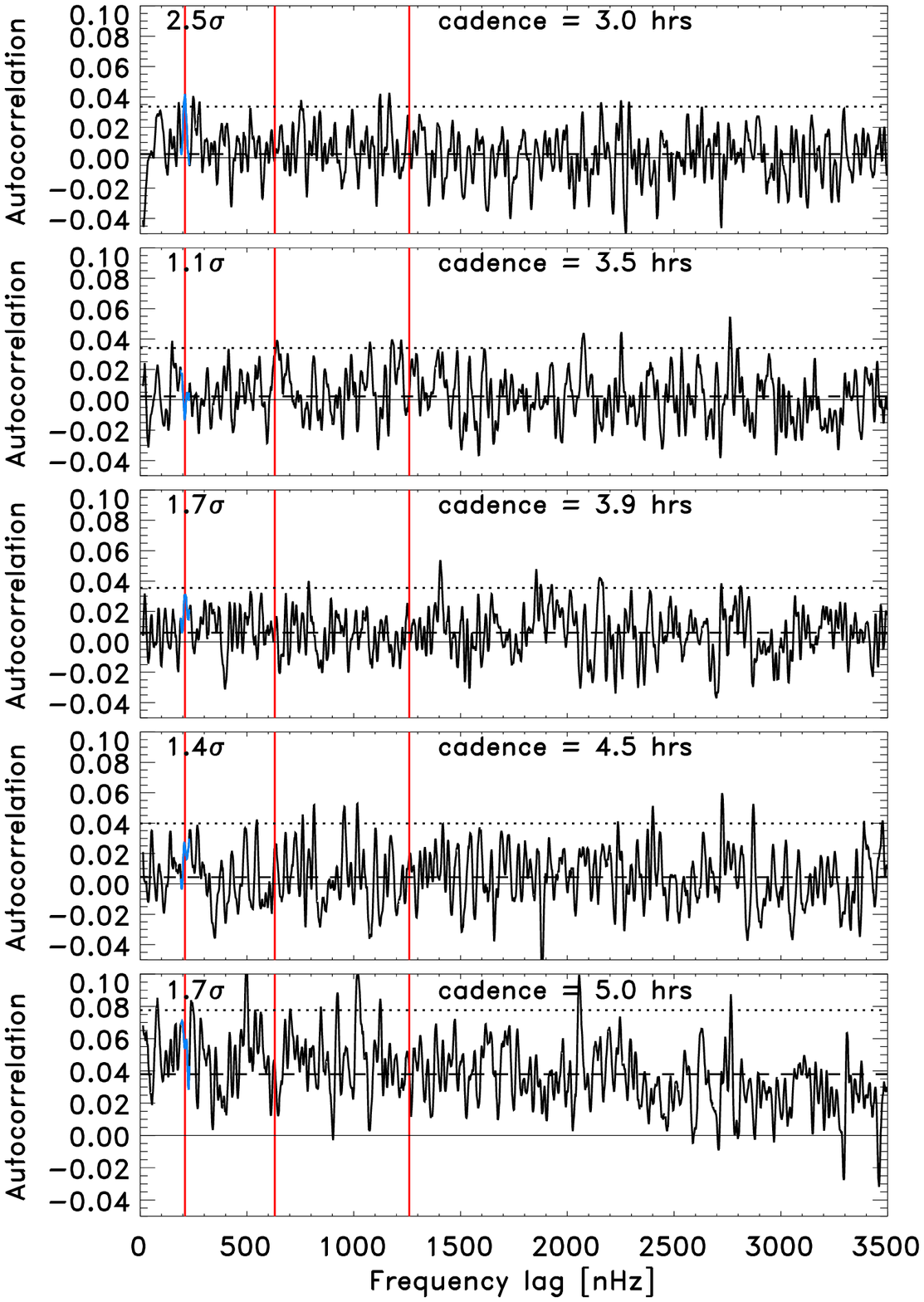}
\caption{The autocorrelation function for different cadences of the round-trip travel-time measurement. From the top panel down we show results for a  3~hr cadence,  3.5~hr cadence, 3.9~hr cadence, 4.5~hr cadence and 5~hr cadence. The blue curve indicates the range within which we find the maxima to compute the significance of the peaks. The vertical red lines indicate where Fossat \textit{et al.} (2017) identified the original three g-mode\ peaks. The signal of the g-mode splittings has vanished in all cases. 
}
\label{fig:rtttcadence}
\end{figure}

Figure~\ref{fig:rtttcadence} shows that the  signal in the autocorrelation vanishes for a 3.9~hr cadence. Why such a small change has such a large impact is difficult to understand, as there is no particular physical relevance of exactly 4~hours.

To further illustrate the robustness of the results we explored what would happen if we had initially selected a different cadence.
To do this, we chose the largest peak in the autocorrelation between 30 and 500~nHz and used this to compute where we would expect the other two peaks to be.
To calculate where the other peaks should be we used the asymptotic splitting of the g modes, 
$s_{\ell,m}  = m \left[\beta_{\ell} \Omega_\mathrm{g} - \Omega_\mathrm{p} \right]$, 
where 
$\beta_{\ell}= 1 - 1/\left( \ell(\ell+1) \right)$, 
where $\Omega_\mathrm{g}$ 
is the mean rotation rate felt by the g modes, and we defined 
$\Omega_p=433$~nHz as the mean rotation rate felt by the p modes. 
For some cases we found peaks where we would expect them to be, albeit with less significance than for the 4~hr cadence case. In other cases we did not find any peaks.

\begin{figure}
\includegraphics[width=0.9\textwidth]{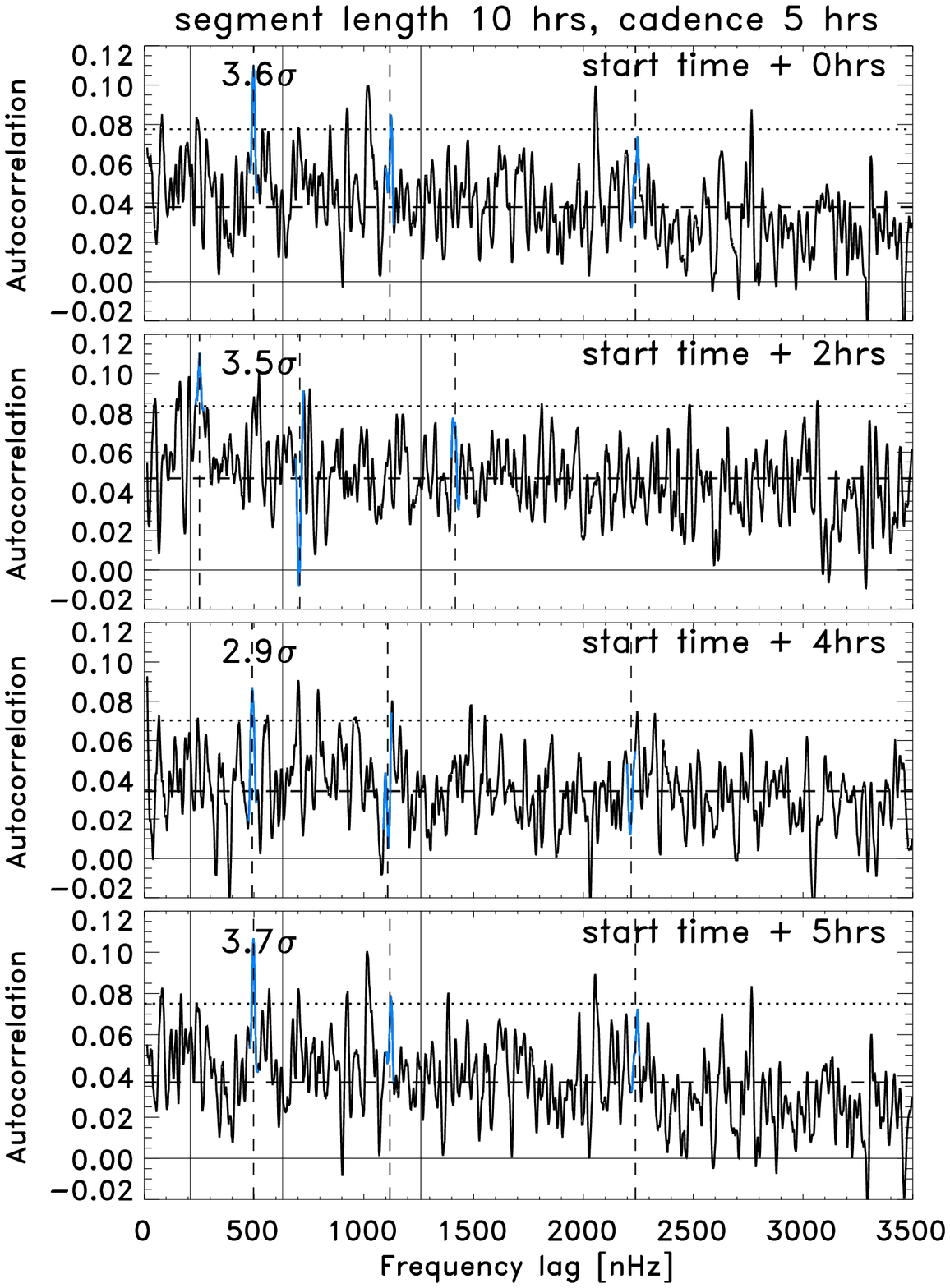}
\caption{The autocorrelation function of round-trip travel-times  at a 5~hr cadence.
The autocorrelation for no start-time offset in the top panel is the same as the bottom panel of Fig.~\ref{fig:rtttcadence}, followed by start-times offset by 2~hrs, 4~hrs and 5~hrs.  
The vertical solid lines indicate where Fossat \textit{et al.} (2017) identified the original three g-mode peaks. 
The vertical dashed lines show the location of the g-mode splittings we identified for zero start time offset at 498~nHz, 1118~nHz, and 2237~nHz.
The blue curve indicates the range within which we find the maxima to compute the significance of the peaks.
}
\label{fig:acchop10}
\end{figure}

As an example, we show the case of measuring the round-trip travel-time at a 5~hr cadence.
We assumed that the largest peak we found, at 498~nHz, is the $\ell=1, m=1$ mode, and so $\Omega_\mathrm{g} = 1862$~nHz (4.3 times the rotation rate felt by the p modes).
Then we used this value of $\Omega_\mathrm{g}$ to predict where the $\ell=2$ mode splittings should be. These values, $s_{2,1}=1118$~nHz and $s_{2,2}=2237$~nHz, are shown as vertical dashed lines in the top panel of Fig.~\ref{fig:acchop10}, and indeed there are peaks nearby. 
This demonstrates that if this arbitrary cadence had been chosen from the beginning, then a different answer would have been found.

\begin{figure}
\includegraphics[width=0.9\textwidth]{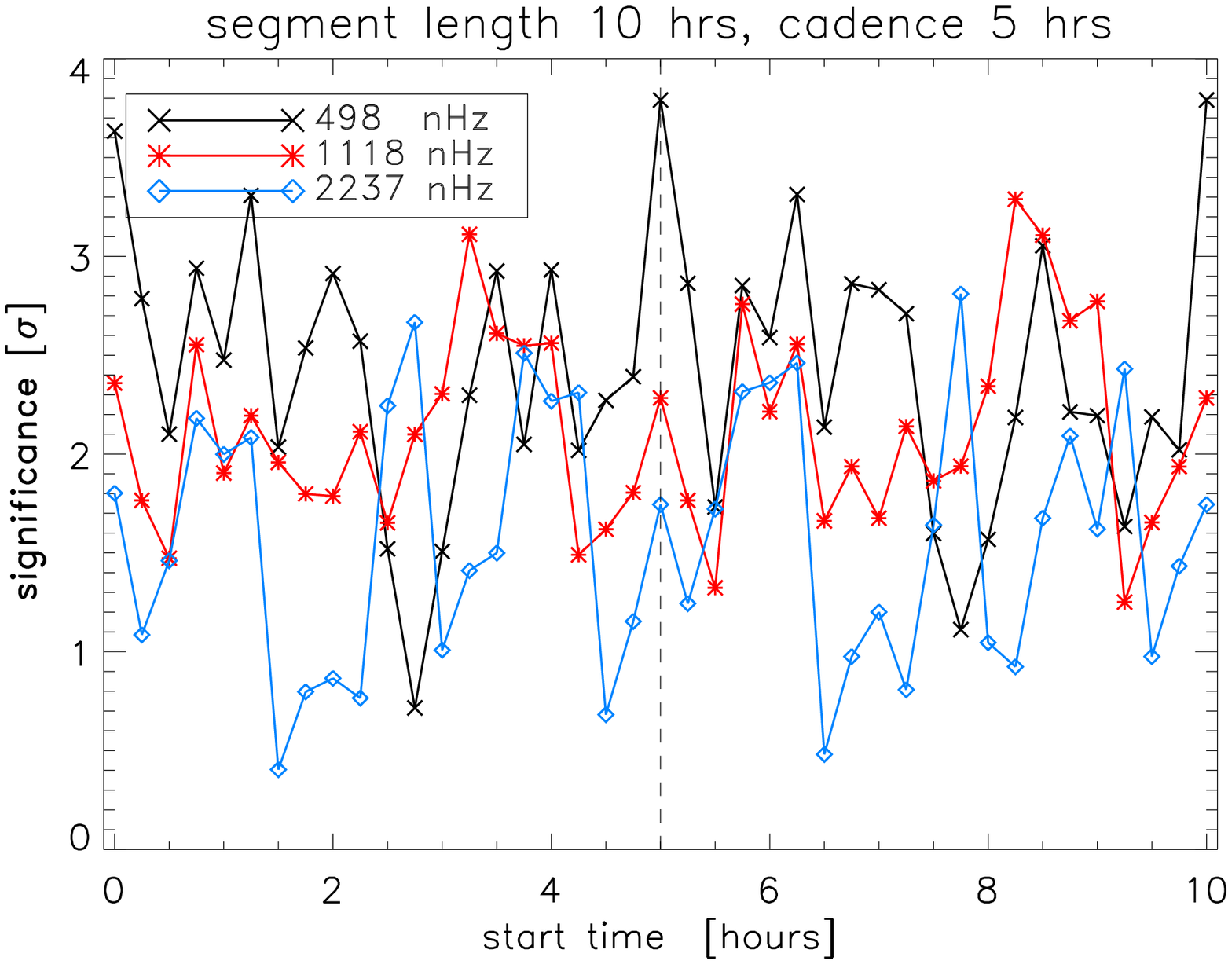}
\caption{The same as Fig.~\ref{fig:acchopfine}, except for segment lengths of 10~hrs and a cadence of 5~hrs.
The points at zero start-time correspond to the three peaks in the top panel of  Fig.~\ref{fig:acchop10}. 
}
\label{fig:ac10chopfine}
\end{figure}

In addition, we adjusted the start time and repeated the analysis for the 5~hr cadence (Fig.~\ref{fig:acchop10}). The bottom panel of Fig.~\ref{fig:acchop10} shows the autocorrelation for a start time offset by 5~hrs (the cadence). We see that the location of the peaks are similar to the case with no offset (top panel). Similarly to the 4~hr cadence case, Fig.~\ref{fig:ac10chopfine} shows that the significances are highest when removing multiples of the cadence.
Fig.~\ref{fig:acchopfine} and \ref{fig:ac10chopfine} together suggest that the peaks found by \citet{Fossatetal2017} could be an artefact of the analysis method and the choice of cadence.

\section{Conclusions}\label{sect:conc}
We have shown that the most recent detection of g modes by \citet{Fossatetal2017} is extremely fragile. 
In particular, the claimed detection is sensitive to 
\begin{itemize}
\item the start-time of the GOLF data series (Fig.~\ref{fig:acchop}), 
\item the cadence of the round-trip travel-time measurements (Fig.~\ref{fig:rtttcadence})
\item the technique used to measure the round-trip travel times (Fig.~\ref{fig:acfits}), and
\item the smoothing of the power spectrum (Fig.~\ref{fig:acfitsmoothing}). 
\end{itemize} 
The first point is particularly worrying because it means that the inclusion or exclusion of a very small fraction of the data makes a substantial difference, and in general because the parameter is unrelated to the properties of the Sun.
This is further illustrated in Fig.~\ref{fig:acchopfine}, which shows that the significance varies with a period equal to the cadence of the data segments used.
The latter two points, on the other hand, are less worrying as they could in principle be attributed to properties of g-mode oscillations, even if the reason for the high sensitivity is not understood in detail.
Overall, we conclude that the  claimed detection of g modes must be treated with extreme caution until these issues are understood. 

It would be valuable to repeat the analysis using other observational data sets from space (MDI, HMI) and from the ground (BiSON, GONG). 

We note that we have not investigated the further analysis by \citet{Fossatetal2017} and  \citet{Fossatetal2018} where they identify a model of the g-mode power spectrum in the observed power spectrum.


\acknowledgements
We thank D.~Salabert and T.~Corbard for helping to interpret the phrase  ``6-bin smoothing" in the \citet{Fossatetal2017} paper, and E.~Fossat for open and helpful discussions.
This work was supported in part by the German space agency (Deutsches Zentrum f\"ur Luft- und Raumfahrt) under PLATO grant 50OO1501.

\bibliographystyle{spr-mp-sola}

\bibliography{papers_bib}

\begin{thebibliography}{6}
\ifx\bisbn     \undefined \def\bisbn  #1{ISBN #1}\fi
\ifx\binits    \undefined \def\binits#1{#1}\fi
\ifx\bauthor   \undefined \def\bauthor#1{#1}\fi
\ifx\batitle   \undefined \def\batitle#1{#1}\fi
\ifx\bjtitle   \undefined \def\bjtitle#1{\textit{#1}}\fi
\ifx\bvolume   \undefined \def\bvolume#1{\textbf{#1}}\fi
\ifx\byear     \undefined \def\byear#1{#1}\fi
\ifx\bissue    \undefined \def\bissue#1{#1}\fi
\ifx\bfpage    \undefined \def\bfpage#1{#1}\fi
\ifx\blpage    \undefined \def\blpage #1{#1}\fi
\ifx\burl      \undefined \def\burl#1{\textsf{#1}}\fi
\ifx\href      \undefined \def\href#1#2{\textsf{#2}}\fi
\ifx\betal     \undefined \def\betal{\textit{et al.}}\fi
\ifx\bctitle   \undefined \def\bctitle#1{#1}\fi
\ifx\beditor   \undefined \def\beditor#1{#1}\fi
\ifx\bbtitle   \undefined \def\bbtitle#1{\textit{#1}}\fi
\ifx\bedition  \undefined \def\bedition#1{#1}\fi
\ifx\bseriesno \undefined \def\bseriesno#1{\textbf{#1}}\fi
\ifx\blocation \undefined \def\blocation#1{#1}\fi
\ifx\bsertitle \undefined \def\bsertitle#1{\textit{#1}}\fi
\ifx\bsnm      \undefined \def\bsnm#1{#1}\fi
\ifx\bsuffix   \undefined \def\bsuffix#1{#1}\fi
\ifx\bparticle \undefined \def\bparticle#1{#1}\fi
\ifx\barticle  \undefined \def\barticle#1{}\fi
\ifx\binstitute  \undefined \def\binstitute#1{#1}\fi
\ifx\bpublisher  \undefined \def\bpublisher#1{#1}\fi
\ifx\doiurl    \undefined
  \def\doiurl#1{\href{http://dx.doi.org/#1}{\textsf{DOI}}}\fi
\ifx\arxivurl  \undefined
  \def\arxivurl#1{\href{http://arxiv.org/abs/#1}{\textsf{arXiv}}}\fi
\ifx\adsurl    \undefined
  \def\adsurl#1{\href{http://adsabs.harvard.edu/abs/#1}{\textsf{ADS}}}\fi
\ifx\botherref \undefined \def\botherref#1{}\fi
\ifx\url       \undefined \def\url#1{\textsf{#1}}\fi
\ifx\bchapter  \undefined \def\bchapter#1{}\fi
\ifx\bbook     \undefined \def\bbook#1{}\fi
\ifx\bcomment  \undefined \def\bcomment#1{#1}\fi
\ifx\oauthor   \undefined \def\oauthor#1{#1}\fi
\ifx\citeauthoryear \undefined\def \citeauthoryear#1{#1}\fi
\ifx\endbibitem\undefined \def\endbibitem{}\fi
\ifx\bconflocation  \undefined \def\bconflocation#1{#1} \fi

\bibitem[\protect\citeauthoryear{{Appourchaux}
  \textit{et~al.}}{2010}]{Appourchauxetal2010}
\begin{barticle}
\bauthor{\bsnm{{Appourchaux}}, \binits{T.}},
\bauthor{\bsnm{{Belkacem}}, \binits{K.}},
\bauthor{\bsnm{{Broomhall}}, \binits{A.-M.}},
\bauthor{\bsnm{{Chaplin}}, \binits{W.J.}},
\bauthor{\bsnm{{Gough}}, \binits{D.O.}},
\bauthor{\bsnm{{Houdek}}, \binits{G.}},
\bauthor{\bsnm{{Provost}}, \binits{J.}},
\bauthor{\bsnm{{Baudin}}, \binits{F.}},
\bauthor{\bsnm{{Boumier}}, \binits{P.}},
\bauthor{\bsnm{{Elsworth}}, \binits{Y.}},
\bauthor{\bsnm{{Garc{\'{\i}}a}}, \binits{R.A.}},
\bauthor{\bsnm{{Andersen}}, \binits{B.N.}},
\bauthor{\bsnm{{Finsterle}}, \binits{W.}},
\bauthor{\bsnm{{Fr{\"o}hlich}}, \binits{C.}},
\bauthor{\bsnm{{Gabriel}}, \binits{A.}},
\bauthor{\bsnm{{Grec}}, \binits{G.}},
\bauthor{\bsnm{{Jim{\'e}nez}}, \binits{A.}},
\bauthor{\bsnm{{Kosovichev}}, \binits{A.}},
\bauthor{\bsnm{{Sekii}}, \binits{T.}},
\bauthor{\bsnm{{Toutain}}, \binits{T.}},
\bauthor{\bsnm{{Turck-Chi{\`e}ze}}, \binits{S.}}:
\byear{2010},
\batitle{{The quest for the solar g modes}}.
\bjtitle{\aapr}
\bvolume{18},
\bfpage{197}.
\doiurl{10.1007/s00159-009-0027-z}.
\end{barticle}
\endbibitem

\bibitem[\protect\citeauthoryear{{Fossat} and
  {Schmider}}{2018}]{Fossatetal2018}
\begin{barticle}
\bauthor{\bsnm{{Fossat}}, \binits{E.}},
\bauthor{\bsnm{{Schmider}}, \binits{F.X.}}:
\byear{2018},
\batitle{{More about solar g modes}}.
\bjtitle{\aap}
\bvolume{612},
\bfpage{L1}.
\doiurl{10.1051/0004-6361/201832626}.
\end{barticle}
\endbibitem

\bibitem[\protect\citeauthoryear{{Fossat}
  \textit{et~al.}}{2017}]{Fossatetal2017}
\begin{barticle}
\bauthor{\bsnm{{Fossat}}, \binits{E.}},
\bauthor{\bsnm{{Boumier}}, \binits{P.}},
\bauthor{\bsnm{{Corbard}}, \binits{T.}},
\bauthor{\bsnm{{Provost}}, \binits{J.}},
\bauthor{\bsnm{{Salabert}}, \binits{D.}},
\bauthor{\bsnm{{Schmider}}, \binits{F.X.}},
\bauthor{\bsnm{{Gabriel}}, \binits{A.H.}},
\bauthor{\bsnm{{Grec}}, \binits{G.}},
\bauthor{\bsnm{{Renaud}}, \binits{C.}},
\bauthor{\bsnm{{Robillot}}, \binits{J.M.}},
\bauthor{\bsnm{{Roca-Cort{\'e}s}}, \binits{T.}},
\bauthor{\bsnm{{Turck-Chi{\`e}ze}}, \binits{S.}},
\bauthor{\bsnm{{Ulrich}}, \binits{R.K.}},
\bauthor{\bsnm{{Lazrek}}, \binits{M.}}:
\byear{2017},
\batitle{{Asymptotic g modes: Evidence for a rapid rotation of the solar
  core}}.
\bjtitle{\aap}
\bvolume{604},
\bfpage{A40}.
\doiurl{10.1051/0004-6361/201730460}.
\end{barticle}
\endbibitem

\bibitem[\protect\citeauthoryear{{Gabriel} \textit{et~al.}}{1995}]{GOLF1995}
\begin{barticle}
\bauthor{\bsnm{{Gabriel}}, \binits{A.H.}},
\bauthor{\bsnm{{Grec}}, \binits{G.}},
\bauthor{\bsnm{{Charra}}, \binits{J.}},
\bauthor{\bsnm{{Robillot}}, \binits{J.-M.}},
\bauthor{\bsnm{{Roca Cort{\'e}s}}, \binits{T.}},
\bauthor{\bsnm{{Turck-Chi{\`e}ze}}, \binits{S.}},
\bauthor{\bsnm{{Bocchia}}, \binits{R.}},
\bauthor{\bsnm{{Boumier}}, \binits{P.}},
\bauthor{\bsnm{{Cantin}}, \binits{M.}},
\bauthor{\bsnm{{Cesp{\'e}des}}, \binits{E.}},
\bauthor{\bsnm{{Cougrand}}, \binits{B.}},
\bauthor{\bsnm{{Cr{\'e}tolle}}, \binits{J.}},
\bauthor{\bsnm{{Dam{\'e}}}, \binits{L.}},
\bauthor{\bsnm{{Decaudin}}, \binits{M.}},
\bauthor{\bsnm{{Delache}}, \binits{P.}},
\bauthor{\bsnm{{Denis}}, \binits{N.}},
\bauthor{\bsnm{{Duc}}, \binits{R.}},
\bauthor{\bsnm{{Dzitko}}, \binits{H.}},
\bauthor{\bsnm{{Fossat}}, \binits{E.}},
\bauthor{\bsnm{{Fourmond}}, \binits{J.-J.}},
\bauthor{\bsnm{{Garc{\'{\i}}a}}, \binits{R.A.}},
\bauthor{\bsnm{{Gough}}, \binits{D.}},
\bauthor{\bsnm{{Grivel}}, \binits{C.}},
\bauthor{\bsnm{{Herreros}}, \binits{J.M.}},
\bauthor{\bsnm{{Lagard{\`e}re}}, \binits{H.}},
\bauthor{\bsnm{{Moalic}}, \binits{J.-P.}},
\bauthor{\bsnm{{Pall{\'e}}}, \binits{P.L.}},
\bauthor{\bsnm{{P{\'e}trou}}, \binits{N.}},
\bauthor{\bsnm{{Sanchez}}, \binits{M.}},
\bauthor{\bsnm{{Ulrich}}, \binits{R.}},
\bauthor{\bsnm{{van der Raay}}, \binits{H.B.}}:
\byear{1995},
\batitle{{Global Oscillations at Low Frequency from the SOHO Mission (GOLF)}}.
\bjtitle{\solphys}
\bvolume{162},
\bfpage{61}.
\doiurl{10.1007/BF00733427}.
\end{barticle}
\endbibitem

\bibitem[\protect\citeauthoryear{{Garc{\'{\i}}a}
  \textit{et~al.}}{2007}]{Garciaetal2007}
\begin{barticle}
\bauthor{\bsnm{{Garc{\'{\i}}a}}, \binits{R.A.}},
\bauthor{\bsnm{{Turck-Chi{\`e}ze}}, \binits{S.}},
\bauthor{\bsnm{{Jim{\'e}nez-Reyes}}, \binits{S.J.}},
\bauthor{\bsnm{{Ballot}}, \binits{J.}},
\bauthor{\bsnm{{Pall{\'e}}}, \binits{P.L.}},
\bauthor{\bsnm{{Eff-Darwich}}, \binits{A.}},
\bauthor{\bsnm{{Mathur}}, \binits{S.}},
\bauthor{\bsnm{{Provost}}, \binits{J.}}:
\byear{2007},
\batitle{{Tracking Solar Gravity Modes: The Dynamics of the Solar Core}}.
\bjtitle{Science}
\bvolume{316},
\bfpage{1591}.
\doiurl{10.1126/science.1140598}.
\end{barticle}
\endbibitem

\bibitem[\protect\citeauthoryear{{Rauer} \textit{et~al.}}{2014}]{PLATO2014}
\begin{barticle}
\bauthor{\bsnm{{Rauer}}, \binits{H.}},
\bauthor{\bsnm{{Catala}}, \binits{C.}},
\bauthor{\bsnm{{Aerts}}, \binits{C.}},
\bauthor{\bsnm{{Appourchaux}}, \binits{T.}},
\bauthor{\bsnm{{Benz}}, \binits{W.}},
\bauthor{\bsnm{{Brandeker}}, \binits{A.}},
\bauthor{\bsnm{{Christensen-Dalsgaard}}, \binits{J.}},
\bauthor{\bsnm{{Deleuil}}, \binits{M.}},
\bauthor{\bsnm{{Gizon}}, \binits{L.}},
\bauthor{\bsnm{{Goupil}}, \binits{M.-J.}},
\bauthor{\bsnm{{G{\"u}del}}, \binits{M.}},
\bauthor{\bsnm{{Janot-Pacheco}}, \binits{E.}},
\bauthor{\bsnm{{Mas-Hesse}}, \binits{M.}},
\bauthor{\bsnm{{Pagano}}, \binits{I.}},
\bauthor{\bsnm{{Piotto}}, \binits{G.}},
\bauthor{\bsnm{{Pollacco}}, \binits{D.}},
\bauthor{\bsnm{{Santos}}, \binits{{\. C}.}},
\bauthor{\bsnm{{Smith}}, \binits{A.}},
\bauthor{\bsnm{{Su{\'a}rez}}, \binits{J.-C.}},
\bauthor{\bsnm{{Szab{\'o}}}, \binits{R.}},
\bauthor{\bsnm{{Udry}}, \binits{S.}},
\bauthor{\bsnm{{Adibekyan}}, \binits{V.}},
\bauthor{\bsnm{{Alibert}}, \binits{Y.}},
\bauthor{\bsnm{{Almenara}}, \binits{J.-M.}},
\bauthor{\bsnm{{Amaro-Seoane}}, \binits{P.}},
\bauthor{\bsnm{{Eiff}}, \binits{M.A.-v.}},
\bauthor{\bsnm{{Asplund}}, \binits{M.}},
\bauthor{\bsnm{{Antonello}}, \binits{E.}},
\bauthor{\bsnm{{Barnes}}, \binits{S.}},
\bauthor{\bsnm{{Baudin}}, \binits{F.}},
\bauthor{\bsnm{{Belkacem}}, \binits{K.}},
\bauthor{\bsnm{{Bergemann}}, \binits{M.}},
\bauthor{\bsnm{{Bihain}}, \binits{G.}},
\bauthor{\bsnm{{Birch}}, \binits{A.C.}},
\bauthor{\bsnm{{Bonfils}}, \binits{X.}},
\bauthor{\bsnm{{Boisse}}, \binits{I.}},
\bauthor{\bsnm{{Bonomo}}, \binits{A.S.}},
\bauthor{\bsnm{{Borsa}}, \binits{F.}},
\bauthor{\bsnm{{Brand{\~a}o}}, \binits{I.M.}},
\bauthor{\bsnm{{Brocato}}, \binits{E.}},
\bauthor{\bsnm{{Brun}}, \binits{S.}},
\bauthor{\bsnm{{Burleigh}}, \binits{M.}},
\bauthor{\bsnm{{Burston}}, \binits{R.}},
\bauthor{\bsnm{{Cabrera}}, \binits{J.}},
\bauthor{\bsnm{{Cassisi}}, \binits{S.}},
\bauthor{\bsnm{{Chaplin}}, \binits{W.}},
\bauthor{\bsnm{{Charpinet}}, \binits{S.}},
\bauthor{\bsnm{{Chiappini}}, \binits{C.}},
\bauthor{\bsnm{{Church}}, \binits{R.P.}},
\bauthor{\bsnm{{Csizmadia}}, \binits{S.}},
\bauthor{\bsnm{{Cunha}}, \binits{M.}},
\bauthor{\bsnm{{Damasso}}, \binits{M.}},
\bauthor{\bsnm{{Davies}}, \binits{M.B.}},
\bauthor{\bsnm{{Deeg}}, \binits{H.J.}},
\bauthor{\bsnm{{D{\'{\i}}az}}, \binits{R.F.}},
\bauthor{\bsnm{{Dreizler}}, \binits{S.}},
\bauthor{\bsnm{{Dreyer}}, \binits{C.}},
\bauthor{\bsnm{{Eggenberger}}, \binits{P.}},
\bauthor{\bsnm{{Ehrenreich}}, \binits{D.}},
\bauthor{\bsnm{{Eigm{\"u}ller}}, \binits{P.}},
\bauthor{\bsnm{{Erikson}}, \binits{A.}},
\bauthor{\bsnm{{Farmer}}, \binits{R.}},
\bauthor{\bsnm{{Feltzing}}, \binits{S.}},
\bauthor{\bsnm{{de Oliveira Fialho}}, \binits{F.}},
\bauthor{\bsnm{{Figueira}}, \binits{P.}},
\bauthor{\bsnm{{Forveille}}, \binits{T.}},
\bauthor{\bsnm{{Fridlund}}, \binits{M.}},
\bauthor{\bsnm{{Garc{\'{\i}}a}}, \binits{R.A.}},
\bauthor{\bsnm{{Giommi}}, \binits{P.}},
\bauthor{\bsnm{{Giuffrida}}, \binits{G.}},
\bauthor{\bsnm{{Godolt}}, \binits{M.}},
\bauthor{\bsnm{{Gomes da Silva}}, \binits{J.}},
\bauthor{\bsnm{{Granzer}}, \binits{T.}},
\bauthor{\bsnm{{Grenfell}}, \binits{J.L.}},
\bauthor{\bsnm{{Grotsch-Noels}}, \binits{A.}},
\bauthor{\bsnm{{G{\"u}nther}}, \binits{E.}},
\bauthor{\bsnm{{Haswell}}, \binits{C.A.}},
\bauthor{\bsnm{{Hatzes}}, \binits{A.P.}},
\bauthor{\bsnm{{H{\'e}brard}}, \binits{G.}},
\bauthor{\bsnm{{Hekker}}, \binits{S.}},
\bauthor{\bsnm{{Helled}}, \binits{R.}},
\bauthor{\bsnm{{Heng}}, \binits{K.}},
\bauthor{\bsnm{{Jenkins}}, \binits{J.M.}},
\bauthor{\bsnm{{Johansen}}, \binits{A.}},
\bauthor{\bsnm{{Khodachenko}}, \binits{M.L.}},
\bauthor{\bsnm{{Kislyakova}}, \binits{K.G.}},
\bauthor{\bsnm{{Kley}}, \binits{W.}},
\bauthor{\bsnm{{Kolb}}, \binits{U.}},
\bauthor{\bsnm{{Krivova}}, \binits{N.}},
\bauthor{\bsnm{{Kupka}}, \binits{F.}},
\bauthor{\bsnm{{Lammer}}, \binits{H.}},
\bauthor{\bsnm{{Lanza}}, \binits{A.F.}},
\bauthor{\bsnm{{Lebreton}}, \binits{Y.}},
\bauthor{\bsnm{{Magrin}}, \binits{D.}},
\bauthor{\bsnm{{Marcos-Arenal}}, \binits{P.}},
\bauthor{\bsnm{{Marrese}}, \binits{P.M.}},
\bauthor{\bsnm{{Marques}}, \binits{J.P.}},
\bauthor{\bsnm{{Martins}}, \binits{J.}},
\bauthor{\bsnm{{Mathis}}, \binits{S.}},
\bauthor{\bsnm{{Mathur}}, \binits{S.}},
\bauthor{\bsnm{{Messina}}, \binits{S.}},
\bauthor{\bsnm{{Miglio}}, \binits{A.}},
\bauthor{\bsnm{{Montalban}}, \binits{J.}},
\bauthor{\bsnm{{Montalto}}, \binits{M.}},
\bauthor{\bsnm{{Monteiro}}, \binits{M.J.P.F.G.}},
\bauthor{\bsnm{{Moradi}}, \binits{H.}},
\bauthor{\bsnm{{Moravveji}}, \binits{E.}},
\bauthor{\bsnm{{Mordasini}}, \binits{C.}},
\bauthor{\bsnm{{Morel}}, \binits{T.}},
\bauthor{\bsnm{{Mortier}}, \binits{A.}},
\bauthor{\bsnm{{Nascimbeni}}, \binits{V.}},
\bauthor{\bsnm{{Nelson}}, \binits{R.P.}},
\bauthor{\bsnm{{Nielsen}}, \binits{M.B.}},
\bauthor{\bsnm{{Noack}}, \binits{L.}},
\bauthor{\bsnm{{Norton}}, \binits{A.J.}},
\bauthor{\bsnm{{Ofir}}, \binits{A.}},
\bauthor{\bsnm{{Oshagh}}, \binits{M.}},
\bauthor{\bsnm{{Ouazzani}}, \binits{R.-M.}},
\bauthor{\bsnm{{P{\'a}pics}}, \binits{P.}},
\bauthor{\bsnm{{Parro}}, \binits{V.C.}},
\bauthor{\bsnm{{Petit}}, \binits{P.}},
\bauthor{\bsnm{{Plez}}, \binits{B.}},
\bauthor{\bsnm{{Poretti}}, \binits{E.}},
\bauthor{\bsnm{{Quirrenbach}}, \binits{A.}},
\bauthor{\bsnm{{Ragazzoni}}, \binits{R.}},
\bauthor{\bsnm{{Raimondo}}, \binits{G.}},
\bauthor{\bsnm{{Rainer}}, \binits{M.}},
\bauthor{\bsnm{{Reese}}, \binits{D.R.}},
\bauthor{\bsnm{{Redmer}}, \binits{R.}},
\bauthor{\bsnm{{Reffert}}, \binits{S.}},
\bauthor{\bsnm{{Rojas-Ayala}}, \binits{B.}},
\bauthor{\bsnm{{Roxburgh}}, \binits{I.W.}},
\bauthor{\bsnm{{Salmon}}, \binits{S.}},
\bauthor{\bsnm{{Santerne}}, \binits{A.}},
\bauthor{\bsnm{{Schneider}}, \binits{J.}},
\bauthor{\bsnm{{Schou}}, \binits{J.}},
\bauthor{\bsnm{{Schuh}}, \binits{S.}},
\bauthor{\bsnm{{Schunker}}, \binits{H.}},
\bauthor{\bsnm{{Silva-Valio}}, \binits{A.}},
\bauthor{\bsnm{{Silvotti}}, \binits{R.}},
\bauthor{\bsnm{{Skillen}}, \binits{I.}},
\bauthor{\bsnm{{Snellen}}, \binits{I.}},
\bauthor{\bsnm{{Sohl}}, \binits{F.}},
\bauthor{\bsnm{{Sousa}}, \binits{S.G.}},
\bauthor{\bsnm{{Sozzetti}}, \binits{A.}},
\bauthor{\bsnm{{Stello}}, \binits{D.}},
\bauthor{\bsnm{{Strassmeier}}, \binits{K.G.}},
\bauthor{\bsnm{{{\v S}vanda}}, \binits{M.}},
\bauthor{\bsnm{{Szab{\'o}}}, \binits{G.M.}},
\bauthor{\bsnm{{Tkachenko}}, \binits{A.}},
\bauthor{\bsnm{{Valencia}}, \binits{D.}},
\bauthor{\bsnm{{Van Grootel}}, \binits{V.}},
\bauthor{\bsnm{{Vauclair}}, \binits{S.D.}},
\bauthor{\bsnm{{Ventura}}, \binits{P.}},
\bauthor{\bsnm{{Wagner}}, \binits{F.W.}},
\bauthor{\bsnm{{Walton}}, \binits{N.A.}},
\bauthor{\bsnm{{Weingrill}}, \binits{J.}},
\bauthor{\bsnm{{Werner}}, \binits{S.C.}},
\bauthor{\bsnm{{Wheatley}}, \binits{P.J.}},
\bauthor{\bsnm{{Zwintz}}, \binits{K.}}:
\byear{2014},
\batitle{{The PLATO 2.0 mission}}.
\bjtitle{Experimental Astronomy}
\bvolume{38},
\bfpage{249}.
\doiurl{10.1007/s10686-014-9383-4}.
\end{barticle}
\endbibitem

\end{thebibliography}

\IfFileExists{\jobname.bbl}{} {\typeout{}
\typeout{****************************************************}
\typeout{****************************************************}
\typeout{** Please run "bibtex \jobname" to obtain} \typeout{**
the bibliography and then re-run LaTeX} \typeout{** twice to fix
the references !}
\typeout{****************************************************}
\typeout{****************************************************}
\typeout{}}

\end{article} 
\end{document}